\documentclass[aps,a4paper,showpacs,amsmath,amsfonts,amssymb,,twocolumn]{revtex4}

\usepackage{graphicx}
\usepackage{epsfig}

\def\(({\left(} \def\)){\right)}
\def\[[{\left[} \def\]]{\right]}

\newcommand{\be}{\begin{equation}}
\newcommand{\ee}{\end{equation}}
\newcommand{\bea}{\begin{eqnarray}}
\newcommand{\eea}{\end{eqnarray}}

\newcommand{\atanh}{\text {atanh}}

\begin{document}
\date{\today}

\title{Glassy aspects of melting dynamics \\ {\it On melting dynamics and the glass transition, Part I} }

\author {Florent Krzakala $^{1,2}$ and Lenka Zdeborov\'a $^{2,3}$}
\affiliation{$^1$ CNRS and ESPCI ParisTech, 10 rue Vauquelin, UMR 7083 Gulliver, Paris 75005 France \\
  $^2$ Theoretical Division and Center for Nonlinear Studies, Los
  Alamos National Laboratory, NM 87545 USA\\
  $^3$ Institut de Physique Th\'eorique, CEA/DSM/IPhT-CNRS/URA 2306
  CEA-Saclay, F-91191 Gif-sur-Yvette, France
}

\begin{abstract}
  The following properties are in the present literature associated
  with the behaviour of super-cooled glass-forming liquids: faster
  than exponential growth of the relaxation time, dynamical
  heterogeneities, growing point-to-set correlation length, crossover
  from mean field behaviour to activated dynamics. In this paper we
  argue that these properties are also present in a much simpler
  situation, namely the melting of the bulk of an ordered phase beyond
  a first order phase transition point. This is a promising path
  towards a better theoretical, numerical and experimental
  understanding of the above phenomena and of the physics of
  super-cooled liquids. We discuss in detail the analogies and the
  differences between the glass and the bulk melting transitions.
\end{abstract}


\maketitle

Almost any liquid becomes a glass when cooled fast enough
\cite{GLASS-GEN:1,GLASS-GEN:2,ANGELL}. Many different scenarios and
theories have been proposed over the time to describe the nature of
glasses. Yet, it is still not known what is the fundamental principle
behind the experimentally observed abrupt change in the relaxation
time of super-cooled liquids.  Is there an underlying critical
phenomenon behind the glass transition or not? Is the glass transition
a thermodynamic or purely dynamic notion? What is the correct theory
of super-cooled liquids? These questions remain unanswered and widely
discussed.

It is the experimental and numerical observations on super-cooled
liquids that remind us of thermodynamic phase transitions. The most
remarkable experimental facts about the transition from liquids to
glasses are indeed: (a) The extremely fast rise of the relaxation time
$\tau$, that increases easily by several orders of magnitude as the
temperature is decreased by a few percent. It is well approximated by
the Vogel-Fulcher-Talman law $\tau \propto
\exp{[A/{\((T-T_{VFT}\))}]}$ \cite{VogelFulcher}. (b) The extrapolated
temperature $T_{\rm VFT}$ is found to be very close to the Kauzmann
temperature $T_K$ where the extrapolated entropy of the supercooled
liquid becomes smaller than that of the crystal \cite{Kauzmann}. (c)
The Adam-Gibbs relation $\tau \propto \exp{[C/\Delta S(T)]}$
\cite{AdamGibbs}, where $\Delta S(T)$ is the difference between the
two entropies, is observed with a good accuracy.

It has been suggested that this
quasi-universal behaviour should be related to the existence of a
length scale growing as the glass transition is approached, and the
hunt for such a length has been a central theme in the field in the last decade. A purely
{\it dynamical} quantity, proposed originally in the context of
mean-field spin glasses \cite{Silvio}, uses a four-point density
correlator in both time and space, and has led to the notion of
{\it dynamical heterogeneities} and to the so-called dynamical
susceptibility (usually refereed to as $\chi_4$). Indeed, the growing
of dynamical heterogeneities has been observed in glass formers
\cite{DynBB,DynHS}. By contrast, other groups have considered the
possibility of a {\it static} (thermodynamic) growing length scale and
proposed the point-to-set correlation, that is the correlation
of a sub-system with its frozen boundaries
\cite{BiroliBouchaud,DynamicBethe,PointToSet}. Recent numerical and experimental
works confirmed that indeed there seems to be such a growing {\it
  thermodynamic} length scale in supercooled liquids
\cite{PTS-Exp-Num}.

One of the standard routes to the above phenomenology goes through the
Random First Order Theory (RFOT) \cite{KT,KW1,KW2,KTW,GlassMezardParisi}
according to which a good starting point of the glass phenomenology
are mean-field spin glasses, and in particular the $p$-spin glass
model \cite{REM,P-SPIN,XORSAT,MPV}. Mean-field spin glasses have an
interesting phenomenology: they behave like a liquid/paramagnet at
large temperature, and when cooled down the equilibration time
diverges as a power law at a temperature $T_{\rm MCT}$ and the
relaxation process is described by the Mode-Coupling Theory (MCT)
\cite{MCT,MCT2,BiroliBouchaudMCT,Kurchan,DYNAMIC,AndreaSaddle}. The
true {\it thermodynamic} glass transition, however, arises at
$T_K<T_{\rm MCT}$ in these models. According to RFOT, the correct
theory of the glass transition is the finite dimensional counterpart
of this behaviour. A phenomenology called the {\it the mosaic picture}
\cite{AdamGibbs,KW1,KW2,KTW,BiroliBouchaud} is used in order to explain
that in structural glasses the relaxation time diverges only at
$T_K$. This crossover from the power-law divergence at $T_{\rm MCT}$
(in the mean-field) to the super-exponential divergence at $T_K$ (in
finite dimension) is at the root of the RFOT theory. The validity of
this picture is still, however,  disputed (see for instance
\cite{Langer,BiroliBouchaud09} and references therein).

In this work, we argue that the above phenomena, usually associated in
the present literature with the dynamics of supercooled liquids close
to the glass transition, also appear in a much simpler problem -- the
bulk melting process of the fully ordered phase above an ordinary
first order phase transition. All the following ingredients arise: the
crossover from a power-law divergence of the relaxation time to an
activated dynamics and a Vogel-Fulcher-Talman-like divergence; the
presence of a growing length scale associated with point-to-set
correlations; the divergence of the dynamical susceptibility and the
presence of dynamical heterogeneities; the presence of a plateau in
the dynamical correlation function with increasing life-time as the
transition is approached.

Analogies between the glass transition and the first order phase
transitions are often evoked -- as testified by the very name of the
{\it random first order theory}. We argue that the extend to which
these analogies are valid and useful is larger than previously
anticipated.  The existence of this correspondance calls for more
detailed theoretical, numerical and experimental investigations in
systems with a first order phase transition. It offers a simpler way
to understand some of the aspects and phenomenology of glassy
dynamics, and inversly provides new ways to look at the bulk melting
problem as well.

In a subsequent paper \cite{US-PART-II}, we shall go beyond a simple
analogy and show that in some systems there is an {\it exact} mapping
between the equilibrium glassy dynamics and the melting of ordered
phase. This is true in particular in the mean field $p$-spin
models~\cite{REM,P-SPIN,XORSAT}, and also on the Nishimori line
\cite{Nishimori-Original,Nishimori} in finite dimensional systems.

This paper is organized as follows: In the first section, we
briefly introduce the melting problem in systems with a first order phase transition, and
stress the differences between surface melting and bulk melting. In
the second section, we review the properties of bulk melting above a
first order phase transitions on the mean-field level. As an example
we use the exactly solvable $p$-spin ferromagnet on a fully connected
lattice, that undergoes a first order ferromagnetic phase
transition. In the third section, we move to the finite dimensional
case, and discuss briefly the nucleation and growth arguments to
predict properties of melting in finite-dimensional systems. In the
fourth section, we simulate numerically the Potts model on a two
dimensional lattice to show that indeed most of the phenomenology
discussed in the glass transition in finite dimensional systems arises
generically also during the melting of the ordered phase. Section V
explains two crucial differences between dynamics of super-cooled
liquids close to the glass transition and melting in a generic system
with a first order phase transition. In the subsequent paper
\cite{US-PART-II} we then study systems with a first order transition
where these differences disappear.  We finally review our results and
discuss some criticisms of the RFOT scenario in the light of our
findings.

\section{Bulk melting}
In this section we specify what we mean by melting process above a
first order phase transition, and in particular we emphasize that we
deal here with bulk melting \cite{SurfaceMelting_R1}, instead of the
much studied ---and arguably more practically important--- surface
melting process.

Consider a system with a first order phase transition, for instance the
solid-liquid transition in crystalline structures or in
three-dimensional hard spheres \cite{HardSphere}, or any given
ferromagnetic spin system with a first order transition such as the Ising
model in field or the Potts model in temperature \cite{Wu}. 
Start in
the fully ordered state (the crystal for solids, the perfect
packing for hard spheres, or the completely magnetized state in the
spin models) and suddenly change the pressure/temperature/field in
order to put the system in the liquid/paramagnetic/less ordered phase.
The system will melt. How long does this melting
process take, and what are its properties?  How does a {\it solid}
turn into a {\it liquid}?

The occurrence of a discontinuous transition suggests that melting of
crystalline or solid matter in nature is a first order phase change
that requires a finite latent heat, and a nucleation mechanism. This
is indeed the case provided some precautions are taken. Consider for
instances solids, unlike super-cooling of liquids, super-heating of
crystalline solids is observed to be extremely difficult if not
impossible
\cite{SurfaceMelting_R1,SurfaceMelting_R2,SurfaceMelting_R3} in most
experiments. This peculiarity is due to the fact that melting of a
crystal kept at a homogeneous temperature always begins on its free
surface. Surface melting is indeed the dominant mechanism for melting
of solids in nature, as it is a much faster process than the bulk
melting. It was, however, realized that by suppressing surface melting
\cite{Superheating} super-heating to temperatures well above the
equilibrium melting point could be achieved. In this case, one
recovers the nucleation phenomenology of first order phase
transitions, and this is the kind of melting processes that, as we
will show, displays many analogies with glassy dynamics. Such bulk
melting problems and the study of the super-heated solids have
received a boost of experimental and theoretical studies recently
\cite{Superheating,SuperANDHetero}.

Here, we shall work with very simple models, far from realistic
atomic or sphere systems. Our motivation stems from the
fact that we expect properties of bulk melting to be universal
and to be qualitatively similar in solids, magnetic systems, hard
spheres or polymers with a first order transition. Following the classical mapping between spin systems and lattice gas~\cite{LatticeGas}, we
will concentrate on the first order transition in Ising and Potts
spin models. Since these models are more easily amenable to analytic studies and simulations, this will
allow us to discuss universal behaviour, such as spinodal and nucleation mechanisms. Moreover, in theoretical considerations surface melting can be suppressed by simply using
fixed or periodic boundary conditions. Finally, many
activities in studies of the glass transition have been devoted to spin models, this is hence a natural setting for discussing the analogies between glassy and melting dynamics.

\section{Melting in mean-field systems}
In this section, we start by discussing the bulk melting transition at the
mean-field level using a simple solvable model. An elementary example of a mean-field system with a
first order phase transition is the fully-connected ferromagnetic
$p$-spin Ising model, for $p=3$
\be {\cal H}=- J \sum_{ijk} S_{i} S_{j} S_{k} ,
\label{PSPIN-ferro}
\ee 
where the sum is over all $N^3$ values of indices $i,j,k$. This is
a simple generalization of the Curie-Weiss model.

\begin{figure}[t]
\vspace{-0.1cm}
\hspace{-0.6cm}
\includegraphics[width=9cm]{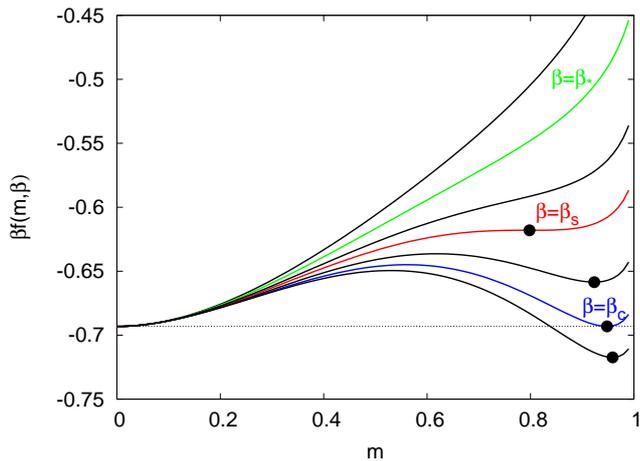}
\caption{(color online): Gibbs free energy $f(m,\beta)$ as a function of the magnetization $m$ for
  different temperatures $\beta$ for the ferromagnetic fully-connected 3-spin model. The figure includes the free energy at the critical temperature $\beta_c=0.67209$ (blue), at the spinodal
  temperature $\beta_s=0.57217$ (red), and at the temperature where the free energy starts to
  be a convex function of the magnetization $\beta_*=0.435$ (green). This illustrates
  the generic behaviour of the free energy in a mean-field system with a first order phase transition.
  \label{FigFIRST-STAT}}
\end{figure}

\subsection{The static behaviour}
Just like the Curie-Weiss model, the ferromagnetic $p$-spin model is
exactly solvable on a fully connected lattice (which is a very crude
mean-field approximation to the finite dimensional case). To ensure
extensivity of the thermodynamic potentials we take
$J=1/N^{p-1}$. The Gibbs free energy as a function of magnetization
$m$ and inverse temperature $\beta$ reads (see the appendix):
\be \beta f(m,\beta)=-\beta m^p-\log{[2 \cosh{\, \atanh{(m)}}]} + {m\, \atanh{(m)}}
\ee while the self-consistent equation for magnetization reads
\be
m = \tanh{\((\beta p m^{p-1}\))} \, .
\ee

As illustrated in Fig.~\ref{FigFIRST-STAT}, for $p=3$, this yields a
first order phase transition at $\beta_c=0.67209$. The low temperature
ferromagnetic phase is, however, locally stable up to the spinodal
point at $\beta_s=0.57217$. The Gibbs free energy $f(m,\beta)$ starts
to be a convex function of magnetization above temperature
corresponding to $\beta_*=0.435$.

\begin{figure}[t]
\hspace{-0.6cm}
\includegraphics[width=9cm]{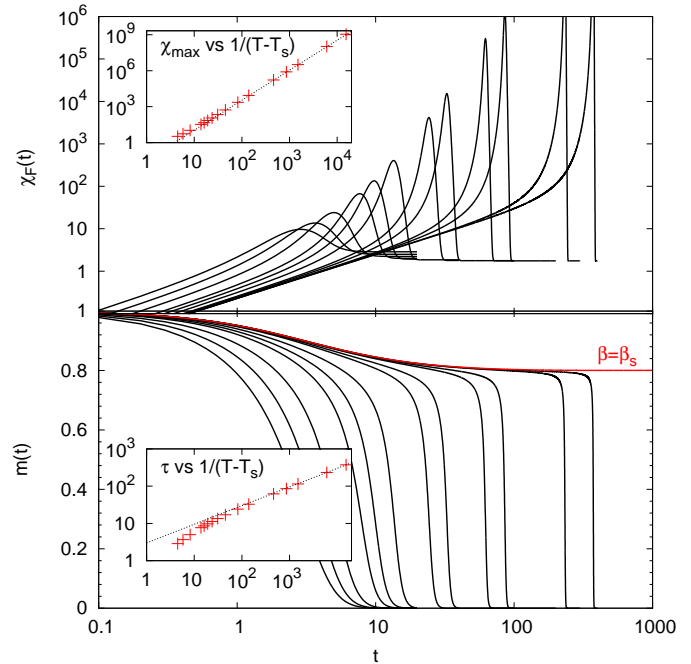}
\caption{(color online) The melting process starting from the
  completely magnetized $m(0)=1$ state for different temperatures in
  the fully connected $3$-spin ferromagnet. Bottom: The time evolution
  of the magnetization $m(t)$, from left to right, $\beta=0.35$,
  $0.4$, $0.45$, $0.5$, $0.52$, $0.54$, $0.56$, $0.565$, $0.57$,
  $0.571$, $0.572$, $0.5721$ and the spinodal temperature (in red)
  $\beta_s=0.57217$. $m(t)$ shows a two-steps relaxation when
  approaching the spinodal point (that starts to be observable at roughly
  $\beta_*$). The time needed for the system to melt diverges
  according to eq.~(\ref{tau}) (see inset). Top: The dynamical
  ferromagnetic susceptibility eq.~(\ref{dyn_susc}) for the same set
  of temperatures. It has a maximum around $\tau$ and converges at
  large times to the static value $\chi_F=\beta$. The peak of the
  susceptibility diverges when approaching the spinodal point as
  eq.~(\ref{chi_max}) (see inset), indicating the growth of a
  correlated volume. \label{FigFIRST-DYN}}
\end{figure}

\subsection{The dynamical behaviour}
Consider the following Glauber dynamics: At each interval of time
$1/N$, we pick up one spin, and set it $+1$ with probability $p_+
=(1+\tanh{\beta h})/2$, and $-1$ with probability $p_- =(1-\tanh{\beta
  h})/2$, where $h=pm^{p-1}$ is the local field on the spin. This
dynamics is also exactly solvable for the fully connected
ferromagnetic $p$-spin model (see again the appendix
for derivations). The average (over
realizations) of the magnetization $\langle m(t)\rangle$ evolves in
time according to the following differential equation
\bea \frac{{\rm d} \langle m(t)\rangle }{{\rm d} t} = -
\langle m(t)\rangle + \tanh{[\beta p \langle m(t)\rangle^{p-1}]}\, .
\label{eq:M-main}
\eea
Fig.~\ref{FigFIRST-DYN} shows how the magnetization evolves in time if
the system is initialized in the ferromagnetic configuration
$m(t=0)=1$ for different temperatures. As the spinodal temperature of the ferromagnetic phase
$\beta_s$ is approached from above, we observe a two-steps
relaxation. The magnetization of the flat region is roughly the
magnetization of the appearing metastable state and its length defines
the relaxation time.  This relaxation time diverges as a power law
\be
\tau \propto \left(1-\frac \beta \beta_s\right)^{-\frac 12}\,
.  \label{tau} \ee
The {\it flattening} of the relaxation process starts roughly at the
temperature $\beta_*$.  The growing plateau in the decay of the order
parameter and the power-law divergence of the relaxation time are
classical features of a melting process above a mean-field first order
transition \cite{Binder73,ReviewBinder}.

\subsection{Static and dynamic diverging length scales}
\label{sec:stat-m}
Coming back to the discussion in the introduction, one may ask if
there is a growing length, or volume, associated with the divergence
of this relaxation time. The naive answer is no: The magnetic
susceptibility that diverges at a second order phase transition stays
indeed finite in systems with a first order phase transition when the ferromagnetic spinodal temperature $T_s$ or the critical temperature $T_c$ is approached from above. This
answer is, however, wrong. Consider the following (time-dependent)
correlation function, averaged over many realizations of the melting
dynamics
\be 
C_{ij}(t) = \langle S_i (t) S_j(t)\rangle  - \langle S_i(t)\rangle \langle S_j(t) \rangle \, .
\ee
It is natural to ask how many spins have correlated moves at a given
time $t$. An estimates of this is given by integrating
this function over the whole system
\bea
\chi_F(t) = \frac{\beta}{N} \sum_{ij} C_{ij}(t)  
= N \beta \left[ \langle m^2(t)\rangle  - \langle m(t)\rangle ^2\right]\, . \label{dyn_susc}
\eea
One recognizes that, for large time, this is nothing but the
equilibrium magnetic susceptibility (we have used the multiplication
by $\beta$ in order to be consistent with the usual definition of the
magnetic susceptibility), and we shall thus call this quantity the
{\it dynamic magnetic susceptibility}. Certainly the equilibrium
magnetic susceptibility is finite in a system with a first order
transition when the ferromagnetic spinodal temperature $T_s$ is
approached from above. Let us look, however, to the finite time
behaviour of $\chi_F(t)$, it follows a differential equation derived in
the appendix. The solution for different temperatures is plotted in
Fig.~\ref{FigFIRST-DYN}.

We see that $\chi_F(t,\beta)$ increases with time, reaches a maximum
roughly at the relaxation time $t=\tau(\beta)$, and then decays to
reach its equilibrium value (which is nothing else than the Curie law
for a paramagnet $\chi_F=\beta$). The maximum value of
$\chi_F(\tau,\beta)$ grows, however, very fast as the spinodal point
$\beta_s$ is approached. The number of spins that are correlated in
the melting process is in this system maximum roughly at the relaxation time
$t=\tau(\beta)$, and thus diverges at the spinodal point as \be
\chi_F(\tau,\beta) \propto \((1-\frac \beta
\beta_s\))^{-\frac{5}{2}}\, . \label{chi_max} \ee This defines the
{\it dynamic} length (or volume) that diverges at the spinodal in the
melting problem. The origin of the critical exponent $5/2$ is
elucidated in a recent work \cite{IwataSasa10}.

\begin{figure}[t]
\vspace{-0.1cm}
\hspace{-0.6cm}
\includegraphics[width=9cm]{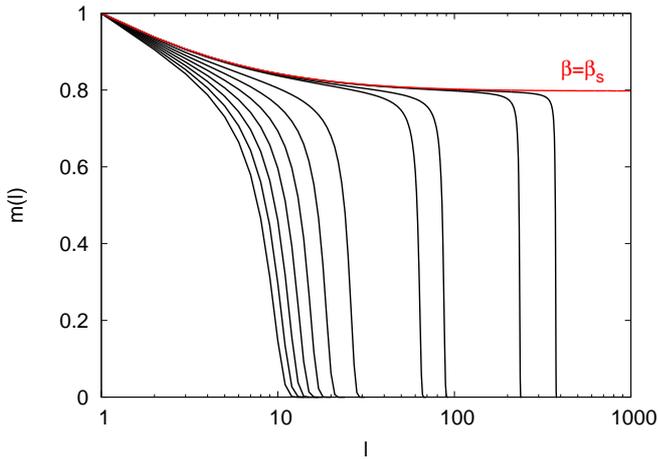}
\caption{(color online) The magnetization $m_l$ at distance $l$ from
  boundaries fixed to positive magnetization in the paramagnetic phase
  of the mean-field ferromagnet. From left to right, we show
  $\beta=0.5, 0.51, 0.52, 0.53, 0.54, 0.55, 0.56, 0.57, 0.571, 0.572,
  0.5721$ and (in red) $\beta_s=0.57217$. The distance one has to go
  from the boundaries in order to recover the equilibrium
  magnetization diverges as eq.~(\ref{pts}) as the spinodal is
  approached. \label{FigFIRST-PTS}}
\end{figure}

Is there also {\it static} volume/length diverging at $\beta_s$?
Again, the answer is positive. In order to see it, we consider the
system with boundary conditions fixed to positive magnetization. In
the paramagnetic phase above the spinodal temperature, fixing the
boundaries in this way does not influence the equilibrium bulk
properties.  But how far should the boundaries be to find back the
zero magnetization phase? This is a well defined point-to-set
correlation length. In a fully connected model it is, however, rather
tricky to define boundary conditions. Let us thus consider the model
on a tree of fixed finite degree. The model can then be solved with
the Bethe-Peierls method (see the appendix). In the large degree limit
the model on tree becomes equivalent to the fully connected one, and
the magnetization at distance $l$ from the boundaries follows the
recursion
\be m_{l+1} = \tanh{\((\beta p m_l^{p-1}\))} \, . 
\label{PTS-M}
\ee 
In Fig.~\ref{FigFIRST-PTS} we show how the magnetization evolves with
distance from the boundaries. One observes that the distances one has
to go in order to recover the zero magnetization diverges when
approaching the spinodal point. Actually, since eq.~(\ref{PTS-M}) is
simply the discrete version of eq.~(\ref{eq:M-main}), this length
diverges as \be \ell \propto \((1-\frac \beta \beta_s\))^{-\frac 12}\,
. \label{pts} \ee

The length $\ell$ we have just defined is in fact well know in the
theory of wetting problems \cite{BOOKWETTING}, where indeed a
divergence is expected. Let us point out that the very same diverging
length on a tree can be defined in another equivalent way: One takes a
{\it finite tree} of depth $\xi$ and computes how large $\xi$ needs
to be such that the magnetization in the center (or the total
magnetization) is bellow a given threshold. On a tree the two
definitions are equivalent and the length $\ell$ is equal to the
length $\xi$ as the two processes are described by the very same
equation. However, the two definitions differ in finite dimensional
systems and we will see that it is the second length $\xi$ that is
associated to the point-to-set correlation length in finite dimension
(while the first one $\ell$ is the finite dimensional wetting length).

Using the fully connected ferromagnet, we have here described the
basic features of the melting process in a mean-field setting. The
time $\tau$ needed to decorrelate diverges as the temperature
approaches the spinodal point. This translates into a growing plateau
in the decay of the order parameter. The divergence of a time scale is
accompanied by at least {\it two diverging length scales}: the {\it
  dynamical} one, that shows how many spins are correlated in the
melting process at the relaxation time scale $\tau$, and a {\it
  static} one, that corresponds to the way the system decorrelates
with ordered boundary conditions.

The learned reader will recognize that these two lengths are nothing
but the dynamical four-point susceptibility and the point-to-set
correlation in disguise. This is the message we want to convey: these
length scales defined initially in the mean field theory of the glass
transition do also diverge in the simpler situation of the melting of
the ordered phase above a first order transition.

At this point, a word of caution should be made: In order to observe
these diverging length and time scales when lowering the temperature
down to the ferromagnetic spinodal $T_s$, it is {\it necessary} to
start from the ordered phase and to consider its melting towards the
high temperature phase. If one instead starts from the equilibrium
liquid/paramagnet state, then the phenomenology is much less
interesting: the {\it equilibrium} relaxation time in the paramagnetic
phase does not show any particular interesting feature, apart from
increasing when approaching the eventual low temperature paramagnetic
spinodal point beyond which the system orders (crystallizes). Of
course, if this ordering is avoided a true glass transition could be
observed at yet lower temperatures, but this is a different phenomena 
from the one described above.

\section{Basics of nucleation theory}
\label{n-and-g}
\label{Sec:3}

In this section, we discuss what happens in finite dimensional systems
and recall briefly the classical nucleation theory in order to explain
that all the diverging time and length scales discussed above should
diverge at the critical point in finite dimensional systems (and not
anymore at the spinodal -- which is no longer thermodynamically well
defined). In the next section, we will illustrate these properties for
the two-dimensional Potts ferromagnet.

In finite dimensional systems the free energy density is always a
convex function of the magnetization, hence the life-time of any
metastable state is always finite. Physically, this is due to the
nucleation of the equilibrium phase that is absent in mean-field
geometries. Metastability hence becomes a purely dynamical notion and
for this very reason giving a proper theoretical definition of a
metastable state in finite dimension can be tricky
\cite{LangerOlder,CavagnaReview}.

Nucleation theory, introduced originally by Gibbs \cite{Gibbs},
describes the way metastable states decay beyond a first order
critical point. The metastable phase has higher free energy, hence it
must decay by nucleation of droplets of the stable (lower free energy)
phase. Nucleation results from a competition between the bulk free
energy difference and the surface tension between the two phases. When
a system is in a metastable phase the free-energy cost $\Delta
F(\ell)$ of a droplet of size $\ell$ is given by
\be \Delta F(\ell) = - V_d \ell^d (\delta f) + S_d \ell^{d-1} \Gamma\, , \ee
where $V_d$ and $S_d$ are the volume and surface factors related to
the shape of the droplet in dimension $d$, $\delta f$ is the
free-energy difference between the two different phases, and $\Gamma$
is the surface tension. The free energy cost is maximum at size
\be \ell_{\rm max} = \frac{(d-1)S_d}{dV_d} \frac{\Gamma}{\delta f} \,
. \label{droplet}\ee
Droplets smaller than $\ell_{\rm max}$ have the tendency to shrink,
while those larger than $\ell_{\rm max}$ will expand fast (the later
process is often described by the Kolmogorov-Johnson-Mehl-Avrami
theory \cite{Avrami}). In order to relax to equilibrium, thermal
fluctuations must cross the free energetic barrier $\Delta F(\ell_{\rm
  max})$. This nucleation time is often estimated by the Arrhenius
formula
\be \tau_{\rm nucl}\approx \exp[\beta \Delta F(\ell_{\rm max})] \approx
\exp\left[\frac{A_d \beta \Gamma^d}{(\delta f)^{d-1}}\right] \, ,
\label{n-time} \ee
where $A_d$ is a dimension dependent constant. The first order melting
point is defined by $\delta f=0$ and therefore both the nucleation time
(\ref{n-time}) and the nucleus size (\ref{droplet}) diverge at the
first-order critical point.

When a critical droplet is formed, it will grow and invade the
system. The relaxation process thus depends on both the
nucleation time {\it and} on the growth of the nucleus. Moreover
relaxation can follow from the appearance of {\it many} droplets in
the system: Indeed the nucleation time $\tau_{\rm nucl}$ gives the
average time needed to nucleate a critical droplet {\it at a given
  place} in the system. If, however, the system is much larger than the
critical nucleus size (a condition that becomes harder and harder to
fulfill as we approach the first-order transition point), the
relaxation process will {\it not} be done via the growth of a single
droplet but instead will result from the nucleation of many droplets
that will simultaneously grow and invade the system. Whether the first
or the second scenario is happening depends on whether the system size
is much larger than the critical nucleus size. In both case, however,
the growth time has to be added to the nucleation time to obtain the
total relaxation time needed to exit from the metastable state.

Consider a large system, much larger than the size of a critical
nuclei. Given the probability to nucleate a critical droplet by a unit
of volume and time is proportional to $\rho \propto e^{-\beta \Delta
  F(l_{\rm max})} \propto 1/\tau_{\rm nucl}$, in a very large system
there should therefore be roughly one droplet per volume $\tau_{\rm
  nucl}$ after a finite time. In most (non disordered) systems, and
for most dynamics, the droplet radius is expected to grow as a power
law $\ell(t) \propto t^{1/z}$ with time. Therefore, after a time of
order $\tau_{\rm nucl}^{z/d}$ droplets will start to touch each other
and the portion of the system in the new phase will percolate. This
predicts that the {\it relaxation time} is $\tau_{relax}\propto
\exp\left[\frac z d \frac{A_d \beta \Gamma^d}{(\delta f)^{d-1}}\right]
$; i.e. it behaves just as the {\it nucleation time}, up to a constant
rescaling of the barrier.  This is the first important conclusion: the
relaxation time diverges super-exponentially at the first order
transition point. Note also that at this time the system is maximally
heterogeneous as a finite portion is the new phase, while the rest is
in the initial one, and we shall see that indeed dynamical
heterogeneities are maximal at this point.

This presentation is oversimplified, as many aspects of the nucleation
and growth process are neglected (for instance the possibility to have
non-compact droplets), and from a generic point of view nucleation
theories could predict more complex exponents in eqs.
(\ref{droplet}-\ref{n-time}), see for instance
\cite{ReviewBinder}. The bottom-line is, however, that when we
consider the melting process in finite dimension, the power-law
divergence at the spinodal point transforms into super-exponentially
slow relaxation at the first order transition
temperature~$T_c$. Consequently, even before the transition the
relaxation time exceeds the experimental time and metastability
becomes observable and unavoidable in practice.  This behaviour of the
relaxation time at a first order phase transition is thus extremely
close to the Vogel-Fulcher phenomenology for glasses
\cite{VogelFulcher}. Moreover, if we consider a first-order phase
transition driven by entropy (as e.g. the $3D$ hard sphere problem)
the free energy difference in eq.~(\ref{n-time}) is replaced by the
entropy difference: in this case the formula looks just like the
Adams-Gibbs relation for glasses~\cite{AdamGibbs}.

\section{Melting dynamics in the 2D Potts ferromagnet}
\label{sec}
We shall now illustrate the above behaviour on the 2D ferromagnetic Potts
model, which is one of the simplest models with a first order phase
transition. It is defined by the Hamiltonian
\be {\cal H} = -\sum_{i,j} \delta(S_i,S_j) \, ,\ee 
where the sum is over all neighboring spins in the two-dimensional
square lattice, and where $S=1,2,\ldots,q$. For $q=2$, it reduces to the Ising model. In two dimensions, one can show \cite{Wu} that
the critical temperature separating the ordered phase from the
paramagnetic one is
\be T_c(q) = \frac 1{\log{(1+\sqrt{q})}}.  \ee
The transition is a first-order one for $q>4$. In order not be too
close to a second order phase transition, we shall work with
$q=10$. In this case, the critical temperature is $T_c(10) \approx
0.70123$. At zero temperature, the ordered configurations are the $10$
ones where all spins take the same value. We shall select one of them,
e.g.  all $S_i=1$, and study the melting of this configuration at
different temperatures. We simulate a lattice of size $L=500$,
i.e. $N=L^2=250000$ spins and use periodic boundary conditions (except
for the point-to-set correlation).

\subsection{Melting and relaxation time}
\begin{figure}[t]
\hspace{-0.6cm}
\includegraphics[width=9cm]{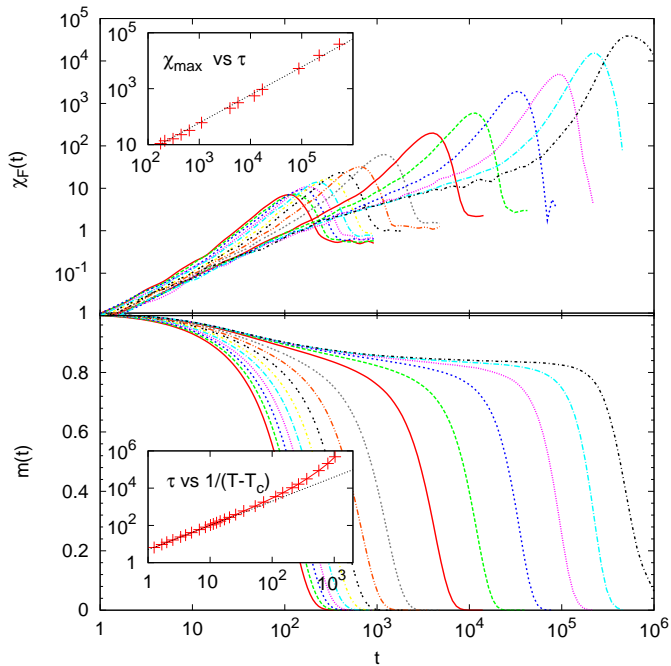}
\caption{(color online): Dynamical susceptibility (top) and decay of
  the magnetization $m(t)$ (bottom) in the melting of the 2D 10-state
  Potts model with $N=250 000$ spins. From left to right: $T=0.8$,
  $0.79$, $0.78$, $0.77$, $0.76$, $0.75$, $0.74$, $0.73$, $0.72$,
  $0.71$, $0.706$, $0.704$, $0.703$, $0.7025$ and $0.7022$. The insets
  show the super-exponential divergence of the relaxation time
  (bottom) and of the maximum susceptibility (top), that is found to
  grow as $\chi_{\rm max} \propto \tau$.\label{dyn-potts}}
\end{figure}

We start by repeating the procedure we used in the mean-field
ferromagnetic model: We initialize the system into a fully ordered
configuration $S_i=1$ and study the dynamics at temperature
$T>T_c$ (for the simulations, we used the heat bath dynamics). 
In Fig.~\ref{dyn-potts} we show the time dependence of the normalized
magnetization
\be m(t) = \frac {q \sum_{i=1}^N [\delta_{S_{i}(t),1}-\frac 1 q]}{N(q-1)}\, . \ee
It develops a plateau when approaching the phase transition, as
typical in glassy systems. The plateau grows super-exponentially fast
when $T \to T_c$, see the inset of Fig.~\ref{dyn-potts}, as predicted
by the nucleation theory. 

\subsection{Point-to-set  like correlations} 
\label{2d-pts}
\begin{figure}
  \vspace{-0.1cm} \hspace{-0.6cm}
\includegraphics[width=9cm]{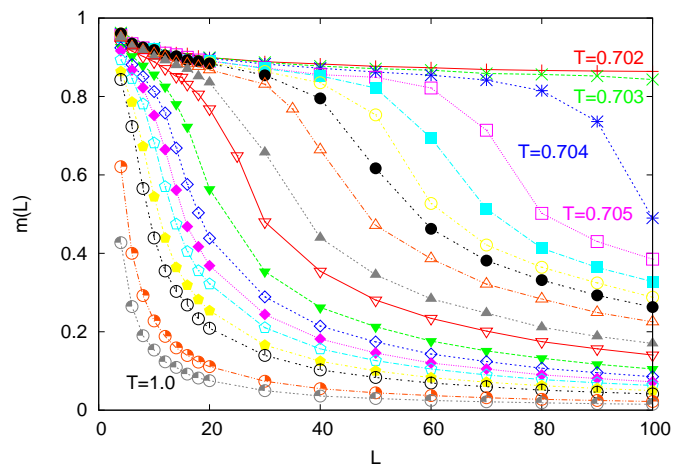}
\caption{(color online) Equilibrium average magnetization in a box of
  size $N=L^2$ with fully magnetized fixed boundary conditions in the
  2D Potts model. These data indicate the growth of the point-to-set
  length scale as the transition at $T_c$ is approached. Defining the
  correlation length $\xi(T)$ such that $m(\xi)<0.8$, we find that $\xi
  \propto (T-T_c)^{-0.8(1)}$, which is reasonably close to the
  prediction $1/(T-T_c)$ of nucleation theory .\label{PTS2d}}
\end{figure}

We shall now fix the boundaries to be in the fully ordered
configuration and study the magnetization they induce for temperature
$T\ge T_c$.  We follow \cite{PTS-Exp-Num} and consider a box of size
$L$ with fixed fully magnetized boundary conditions. We then run a
Monte-Carlo simulation, using the Wolff's cluster algorithm
\cite{Wolff} in order to reach equilibrium\footnote{The algorithm has
  to be slightly modified to take into account the boundaries: instead
  of flipping the cluster with probability $1$, as in \cite{Wolff}, we
  do it with probability $p=\max(1,e^{-\Delta E_b})$ where $E_b$ is
  the energy difference due to the boundaries.}, and measure the total
magnetization inside the systems for different sizes $L$ and
temperatures $T$. The results are shown in Fig.~\ref{PTS2d}. We see
clearly a growing {\it point-to-set} correlation length that diverges
as the transition at $T_c$ is approached. Defining the correlation
length as the moment when the magnetization falls bellow the plateau
(in practice, we used $m<0.8$), we observed that the length grows
approximately as $1/(T-T_c)^{0.8(1)}$, which is not very far from the
linear scaling of the size of a critical nucleus
(\ref{droplet}). Indeed, $\xi(T)$ is nothing but an estimation of the
nucleus average size \cite{Cavagna} in the melting process. As the
temperature drops, melting dynamics is initiated by the reversal of
larger and larger clusters of spins (the nuclei that will later
expand), and the size of these clusters diverges at $T_c$.

A word of caution is needed here: we have identified a diverging
length scale in first-order transition process.  But this length {\it
  not} be confused with the {\it equilibrium correlation length},
which does {\it not} diverge at a first order transition (in fact it
reaches a value of $\approx 7.5$ lattice spacings exactly at the
transition in this model, see \cite{Janke}). The point-to-set length
diverges while the equilibrium correlation one does not.

To echo the discussion in sec. \ref{sec:stat-m}, let us further
emphasize that the point-to-set construction we have used is also
different from taking {\it first} a very large box with ordered
boundary conditions and {\it then} measuring the magnetization at
distance $L$ from the boundaries. Indeed, for box sizes $L \gg \xi$
nucleation is always favorable and the nucleus thus invades almost the
whole system. From the data presented in Fig.~\ref{PTS2d} we can see
that the magnetization for $L \gg \xi$ starts to decay as the inverse
of the size $L$. This can be understood easily: If the correlation
with the boundaries {\it in a very large box} is set by a
characteristic distance $\ell$, then the local magnetization will be
roughly one at distance $d<\ell$ from the boundaries, and roughly zero
at distance $d>\ell$. The total magnetization should thus behave as $m
\propto \ell/L$ when $L\gg \xi$. The characteristic length $\ell$ is,
however, much smaller and unrelated to the point-to-set construction
(in fact, it is in the so-called {\it wetting length}, which is
expected, at least in simple mean field models, to diverge
logarithmically at the transition, see for instance
\cite{BOOKWETTING}).  This is a major difference with the mean field
case where at the spinodal point, both the correlation length with
frozen boundaries in an infinite system (that is, the wetting length)
{\it and} the length defined by the onset of magnetization in a finite
system (that is, the point-to-set length) are equal and diverge in the
very same way. This is not true anymore in finite dimensional systems.

\subsection{Dynamical heterogeneities}
\label{2d-dyn-get}
\begin{figure}[t]
\vspace{-0.3cm}
\hspace{-1cm}
\includegraphics[width=9.0cm]{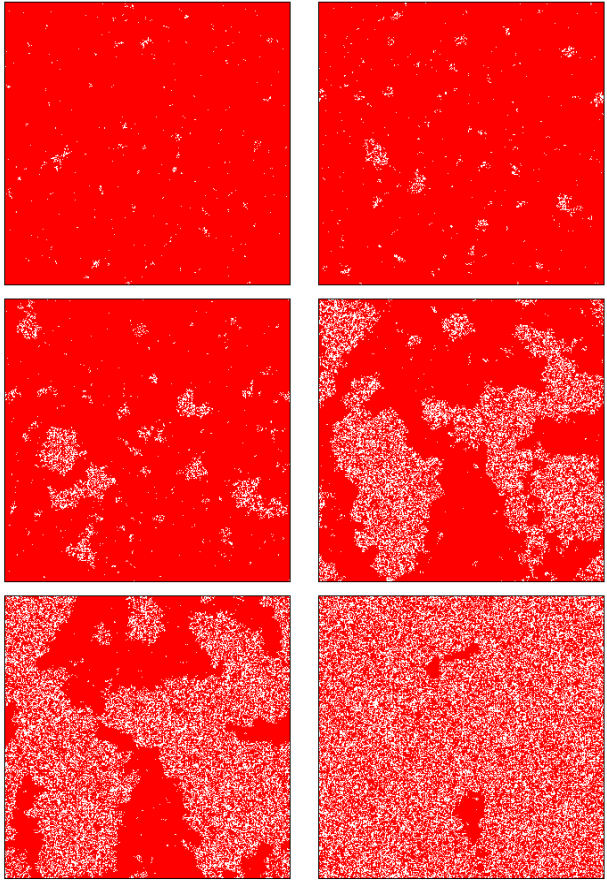}
\caption{(color online) Nucleation process in the melting of the 2D
  Potts model with $q=10$ at temperature $T=0.7025$ with $N=1500^2$
  spins.  The red dots are the spins that have the same value as the
  initial state $S=1$, white are all the other spins. The times are
  from top to bottom $25000$, $75000$, $120000$, $220000$, $260000$,
  $380000$.  On the second image, we can already recognize some
  nucleating critical droplets that start to grow fast for larger
  times. The nucleating droplets percolate between images n.~$4$ and
  n.~$5$, this corresponds to the time at which the heterogeneities
  and the dynamical susceptibility are the largest. The typical size
  of heterogeneities is much larger than the size of critical
  nuclei \label{NUCLEATION}.}
\end{figure}

We also computed the dynamical susceptibility $\chi_F(t)$, see
Fig.~\ref{dyn-potts}. Again, it yields the qualitative behaviour of the
$\chi_4$ observed in glassy systems, meaning that the dynamical
heterogeneities are the largest when the susceptibility reaches its
maximum and they grow as the temperature is approaching the phase
transition.  As we are working with a 2D system, visualization
is convenient and we thus depict these heterogeneities in
Fig.~\ref{NUCLEATION}.

Since the melting proceeds via nucleation, one might think that the
maximum of the $\chi_F(t)$ denotes the volume of the critical droplet
(\ref{droplet}). This is not the case, in fact the volume described by
the maximum of $\chi_F$ is much larger than the critical nucleus size. This
is clearly visible in the simulations depicted in
Fig.~\ref{NUCLEATION}, where the size of a typical critical nucleating
droplet is about $10^2$, while the typical size of the correlated
regions close to the maximum of the $\chi_F$ is about $100^2$.

The reason beyond this is simple and illustrates well the nucleation
and growth mechanism. As described in Sec.~\ref{n-and-g}, after a time
of $O\((\tau_{\rm relax}\))$, grown droplets will touch each other and
percolate; this is the moment at which the maximum of the $\chi_F$
arises, and we expect that this maximum thus behaves algebraically
with the relaxation (and the nucleation) time so that $\chi_F \propto
\tau^{\alpha}$, which is indeed what we find in the inset of
Fig.~\ref{dyn-potts}. This shows that the nature of the static and the
dynamic length scale is completely different, and that the dynamical
susceptibility can probe much larger scales. The first one probes the
length associated with critical nuclei while the second one depends on
the growth process. Note that the exponential divergence of $\chi_F$
(with respect to the temperature difference as the critical point is
approached) compares well with what is observed in many models with
{\it kinetically constraint} or {\it facilitated dynamics}
\cite{Chi4_kinetic_1,Chi4_kinetic_2,Chi4_kinetic_3}.

Given the simplicity of its origin, a law $\chi_F \propto
\tau^{\alpha}$ should indeed be quite generic for systems with
nucleation and growth. However, the value of the exponent
$\alpha$ depends on the type of growth and on the precise properties
of the system. The growth process of nuclei can be in principle much
slower in system with disorder (where, due to the pinning of the
interface, activation is again necessary to grow the nuclei) with
respect to the relatively fast process observed here, and its
interplay with nucleation can be more complicated. In fact, the growth
can probably be so slow that the two length scales might be
comparable, and this will be investigated in a companion paper
\cite{US-PART-II}.

\section{Differences between bulk melting and glassy dynamics}
So far we concentrated on the analogies between glassy dynamics and
bulk melting, it is time to list several key differences. First of
all, the analogy does not extend to non-equilibrium properties such as
{\it aging} in glasses \cite{Kurchan}. We are comparing here the
melting process with the equilibrium-initialized dynamics in the
glassy systems, but the out-of-equilibrium behaviour of glasses is very
different. But let us concentrate on the differences in the dynamics
from equilibrium.

A crucial point is that there is no latent heat associated to the
glass transition, as the energy is continuous, while there {\it is a
  latent heat} in a general first-order transition (for instance both
in the ferromagnetic $p$-spin model and the 2D Potts model we
discussed in this paper).  To smear away this difference one should
consider a first order transition driven purely by entropy, if the
energy is continuous at the first order phase transition then there is
no latent heat. This is the case for instance in a system of hard
spheres.

Another major difference in the dynamical behaviour is that the melting
process happens once for all. When the system has melted, it stays in the
liquid phase. However, the equilibrium glassy dynamics is a stationary
process which is time-translationally invariant.  This is why the
equivalent of the nucleation theory in glasses, the so-called mosaic
picture \cite{AdamGibbs,KW1,KW2,KTW,BiroliBouchaud}, is more
complicated (and not yet fully understood).  

Our results seem, however, to point towards the glass transition being
a melting {\it of some special sort}. In the glass problem, following
the ideas of Goldstein \cite{Goldstein}, we can assume that we have a
more complicated landscape, with many local minima. The glassy
dynamics consists of a continuous and stationary melting of one minima
to another one.  This view is precisely the one adopted in the mosaic
picture and in the RFOT \cite{KW1,KW2,KTW,BiroliBouchaud}.

The comparison between melting and glassy dynamics will be pushed
beyond a simple analogy in a companion paper \cite{US-PART-II}, where
we show that in a certain class of spin systems (both mean-field and
finite dimensional ones) with a first order phase transition, one
observes a purely entropic transition and the melting dynamics is a in
fact equivalent to the equilibrium dynamics. In order to do so we
explore properties on the Nishimori line
\cite{Nishimori-Original,Nishimori,Ozeki1,Ozeki2} and study
consequences of such a mapping for the physics of the glass
transition.

\section{Discussion}
In this paper we explored analogies between the bulk melting dynamics
above a first order phase transition and the equilibrium glassy
dynamics. We showed that many properties associated with the dynamics
of structural glass formers are already present in the melting
case. We have investigated several features of modern theory of
super-cooled liquids --- such as the diverging relaxation time, the
plateau in the correlation function, dynamical heterogeneities related
to the dynamical susceptibility $\chi_4$, the point-to-set
correlations --- and showed that they all can be easily recovered and
understood in a much simpler setting of systems with a first order
phase transition.  In particular the dynamical ---or mode-coupling---
transition corresponds to the spinodal point, while the Kauzmann
transition corresponds to the first order ferromagnetic phase
transition.

Noticing similarities between the glass transition and the first order
one is of course not new. In fact, it is somehow implicit in the
random first order theory of glasses \cite{KW1,KW2,KTW}, where
according to the mosaic picture every glassy state melts into another
one and so on.  The notion of nucleation stands at the roots of the
theory deriving the Adam-Gibbs law and the mosaic theory is nothing
but the counter-part of nucleation theory for glassy transitions. The
construction of Franz and Parisi \cite{FranzParisi} and the earlier
work of \cite{KT} are attempts to root the theory of the glassy
transition into a first-order setting. The recent work of
\cite{CavagnaAgain} also accentuates the analogy, and the very
definition of the point-to-set correlation was recently proposed as a
tool to investigate nucleation in systems with a generic first order
transition \cite{Cavagna}, and this is by far not an exhaustive list.
The bottom-line of our approach here is that it is rather interesting
to take this analogy literally, to concentrate on the simpler melting
problem, and to understand in detail its glassy aspects. We believe
this approach can be interesting for the bulk melting problem {\it per
  se.}

Before concluding, let us make a last comment. The random first order
theory is not fully accepted as a good theory
of the glass transition. Without taking a position on the question whether
RFOT is the good theory for the glass transition or not, we want to
point out here that many of the criticisms of RFOT apply equally to
systems with an ordinary first order phase transition. Imagine for a
moment that we would study melting dynamics without knowing about the
underlying first order phase transition. In this case researchers
could be trying different fits to extrapolate the divergence of the
melting time at a finite or zero temperature
\cite{hecksher2008little}.  They could also criticize the use of
ill-defined metastable states \cite{Langer,Chi4_kinetic_1,Wyart}.
Mean-field theories would predict a spinodal line with a power-law
divergence \cite{MCT,MCT2}, and the crossover to finite dimensional
behaviour could be thought of as a mysterious one
\cite{BiroliBouchaud09}.  There would be some claims that mean-field
theories do not incorporate geometric properties and hence are not
very relevant to finite dimensional systems \cite{Langer}. Some would
concentrate on the heterogeneous fluctuations in the decay of the order
parameter, and on the peculiar shape of critical nucleating droplets
as the central object of interest \cite{PhysRevLett.80.2338}; claiming
at the same time that there is no need to think about thermodynamics
\cite{Ludo} and that melting should be modeled with purely dynamic
systems with simple rules \cite{Chandler}. In cases of systems such as
diamond the melting transition might even be not considered at all
because the true equilibrium state is simple carbon --- just as the crystal
is always the true equilibrium state in glass forming systems
\cite{CRYSTAL}. Luckily, first order phase transitions are much more
established (since in this case, contrary to what happens in glasses,
we {\it can} equilibrate at low temperatures), hence the answers to
the above arguments are mostly well understood. Still, this shows that
starting from a mean-field analysis and correcting it with
nucleation-like arguments, as is done in RFOT, is not such a bad
strategy {\it a priori}.

Finally, a lesson from this study is that one should maybe look back
to the first order phase transitions with the eyes of glassy
phenomenology, both from an experimental and theoretical point of
view. The following questions are for instance appealing: How do the
static and dynamic lengths behave in other models with first order
transitions (with or without disorder)?  What does the mode-coupling
approximation predict for a first order phase transition? Can all the
recent experimental investigations of static and dynamic length scales
in glass formers be repeated in a first order melting process? The way
a solid flows under shear is clearly related to the melting mechanism
\cite{GiulioLast}, does the dynamical susceptibility diverge as a
power of the relaxation time as well?

These questions are particularly appealing in the context of bulk
melting in super-heated solids \cite{Superheating}. It would be for
instance interesting to apply the dynamical correlation and
susceptibility analysis of sections \ref{sec:stat-m} and
\ref{2d-dyn-get} in order to identify dynamical heterogeneities (which
seem indeed to be present in this problem, see for instance
\cite{SuperANDHetero}). This could also help to understand the
interplay between nucleation and growth in melting processes. Other
common behaviour can be identified upon inspection; compare for
instance the {\it string-like} motion \cite{STRING} observed in glassy
dynamics with the highly correlated {\it ring} and {\it loop} atomic
motions behaviour observed in bulk melting \cite{MO}
(and more generally with the stringy nuclei observed close to
pseudo-spinodal during first-order transition \cite{KLEIN}).
 Another example is given by the behaviour of the shear modulus: according to a recent
work within the RFOT, it is predicted to vanish continuously at the
mode-coupling transition point in glasses \cite{MARC}, a statement
clearly reminiscent of the Born criterion \cite{BORN} that defines the
spinodal point of superheated solid as the moment where shear modulus
is going to zero. Finally, studying the system with respect to frozen
boundary conditions as in section \ref{2d-pts} should also be
interesting in order to estimate nucleation barriers numerically in
the spirit of \cite{Cavagna}, and to investigate the much studied
problem of the limit of the super-heated state.

To conclude, there is a lot to learn on the crossroad between glasses
and first order melting, and we hope our article will trigger new
works in this direction. This will be further accentuated in a
companion article where we provide a set of models where the melting
process is {\it exactly} equivalent to its equilibrium dynamics
\cite{US-PART-II} and where we show that the mean-field theory of the
glass transition is mappable to a (special) melting problem.

\acknowledgments It is a pleasure to thank G. Biroli, J-P. Bouchaud,
A. Cavagna, S.~Franz, J. Kurchan, J. Langer and H. Yoshino for
interesting discussions about these issues.

\appendix

\section*{Appendix: The ferromagnetic $p$-spin model}
\label{sec:appendix:ferro}
In this appendix, we remind how to solve the static and the dynamic behaviour of the fully
connected ferromagnetic $p$-spin model defined by Hamiltonian (\ref{PSPIN-ferro}) with $J=1/N^{p-1}$.

\subsection{The thermodynamic solution}
In order to compute the free energy as function of magnetization, we consider the Hamiltonian (\ref{PSPIN-ferro}) with an external magnetic field $h$, it then becomes $H=-N(m^p+hm)$. Applying the integral representation of the delta function, one finds
\bea \nonumber 
 &Z(h)&=
\sum_{\{S_i\}} \int {\rm d}m \, \delta(Nm-M(\{{\vec{S}\}})) \,  e^{ \beta N \[[m^p + h m\]]} \\
&=& \int \!\! {\rm d}m \, {\rm d}\lambda\,  e^{ \beta N (m^p+hm) \! - \! N\!   \lambda m  + N \log{(2 \cosh{\lambda})} } \nonumber.
\eea
The saddle point condition imposes that $\lambda=\beta p m^{p-1}+\beta h$ and therefore
\be
Z(h)=\int {\rm d}m\,  e^{-\beta N g(m,\beta,h)}
\ee
with
\be
g(m,\beta)= (p-1) m^p - \frac 1{\beta} \log{[2 \cosh{\((\beta p m^{p-1} + \beta h \))}]}\, .
\label{fnrg_1}
\ee
The self-consistent equation on $m$ thus reads
\be
m = \tanh{\((\beta p m^{p-1} + \beta h\))} \, .
\label{condition}
\ee
We are interested in the free energy as a function of the equilibrium
magnetization $m$. This is obtained by the Legendre transform of
eq.~(\ref{fnrg_1}), that is, by $f(m,\beta)=g(h^*,\beta,m)+h^*m$, where
$h^*$ is given by the condition (\ref{condition}). It yields finally
\be 
\beta f(m,\beta)=-\beta m^p-\log{\[[2 \cosh{\, (\atanh{\, m})}\]]} +
{m~\atanh(m)}\, . \ee
This formula was used to produce  Fig.~\ref{FigFIRST-STAT}. For $p=3$, this solution yields first-order phase
transition at $\beta_c=0.67209$. The low temperature phase, however,
is locally stable until the spinodal point at $\beta_s=0.57217$.

\subsection{The Bethe-Peierls solution}
Consider the $p$-spin model on a tree with coordination number $c$.
In this case, one can write an iterative recursion using the Bethe-Peierls strategy that relates the
magnetization at level $l+1$ with the one at level $l$ \cite{XORSAT}
\be 
m_{l+1} = \tanh{\left\{ (c-1) {\rm arctanh}{\left[ \tanh{(\beta J)} m_l^{p-1} \right]}\right\}}\, .
\ee
Let us now take the limit of a very large connectivity and use $J=p/c$. In this limit
\be m_{l+1} = \tanh{\(( p \beta m_l^{p-1} \))}\, . \ee
The stationarity condition of this equation leads to
eq.~(\ref{condition}). This formulation allows to define a correlation length as is done in sec.~\ref{sec:stat-m}.

\subsection{Solving the dynamics of the melting process}
We define the Glauber dynamics as follow: At each interval of time
$1/N$, we pick up one spin, and set it plus with probability $p_+
={(1+\tanh{\beta h})}/2$ and minus with probability $p_-
={(1-\tanh{\beta h})}/2$, where $h=pm^{p-1}$ is the local field on the
spin, that is conveniently the same for every spin. To compute the
evolution of magnetization and susceptibility (\ref{dyn_susc}), we
define $P(t,M)$ as the probability (over different realizations of the
dynamics) that the value of the magnetization is $M$ at time $t$,
where $-N \le M \le N$. Its averaged value per site reads 
\be m(t) =
\frac{1}{N} \int_0^1 P(t,M) M {\rm d}m \, , \ee
and the average susceptibility is
 \be \chi(t) = \frac{\beta}{N} \left\{ \int_0^1 P(t,M)
  M^2 {\rm d}m - \left[ \int_0^1 P(t,M) M {\rm d}m \right]^2 \right\}
\, , \ee where $M=Nm$. The master equation for $P(t,M)$ reads
\bea 
&&P(t+{\rm d}t,M) = \frac{N+(M+2)}{2N}\, p_-(M+2)\, P(t,M+2) \nonumber \\
&& + \frac{N-(M-2)}{2N}\, p_+(M-2) \, P(t, M-2) \nonumber \\
&& + \left[\frac{N+M}{2N} p_+(M) + \frac{N-M}{2N} p_-(M)\right] P(t,M)\, .
\label{master} 
\eea 
The average magnetization at time $t+{\rm d}t$ is thus
\be
Nm(t+{\rm d}t)= \int_0^1 P(t,M) {\rm d}m
\left[ M -m + \tanh{(\beta pm^{p-1})} \right] \label{av_m} \, . \ee
Considering that fluctuations in $m$ are $O(1/\sqrt{N})$, this leads to the differential equation (\ref{eq:M-main}).


The evolution of the dynamical magnetic susceptibility is computed in a similar manner. Using (\ref{master}), the second moment of $M$ reads 
\bea
&&\int_0^1 M^2  P(t+{\rm d}t,M)  {\rm d}m = \int_0^1 M^2 P(t,M) {\rm d}m \nonumber \\
 &+& \int_0^1 2 M P(t,M) \left[ -m + \tanh{(\beta pm^{p-1})} \right] {\rm d}m \nonumber \\
&-& \int_0^1 2 P(t,M) \left[ m \tanh{(\beta pm^{p-1})} -1 \right] {\rm  d}m \label{av_m2}\, . \eea
Now from the definition of $\chi(t)$ after collection all the terms
from (\ref{av_m}) and (\ref{av_m2}) we get 
\bea
 &N& \chi(t+{\rm d}t)/\beta = N \chi(t)/\beta \nonumber \\
&+& \int_0^1 2 (M-\langle M \rangle) P(t,M) \left[ -m +
  \tanh{(\beta pm^{p-1})} \right] {\rm d}m \nonumber \\ &+& 2 \langle m
\tanh{(\beta pm^{p-1})} \rangle -2 - \langle -m + \tanh{(\beta
  pm^{p-1})} \rangle^2\, . \nonumber \eea 
The remaining integral on the r.h.s. can be rewritten using a
Taylor series around $\langle m(t) \rangle$. Finally we obtain
the following differential equation for $\chi(t)$
\bea && \frac{{\rm
    d}\chi(t)}{{\rm d}t} = - 2 \chi(t) + 2 \frac{\beta
  p(p-1)m(t)^{p-2}}{\{\cosh{[\beta pm(t)^{p-1}]}\}^2} \chi(t) \nonumber \\
&+& 2\beta -
\beta m^2(t) - \beta \left\{ \tanh{[\beta pm(t)^{p-1}]} \right\}^2\, .
\label{eq:CHI}
\eea
Equations (\ref{eq:M-main}) and (\ref{eq:CHI}) can be numerically
integrated together, and yield the results in Fig.~\ref{FigFIRST-DYN}.

\end{document}